\documentstyle[aps,pra]{revtex}
\begin{document}
\title{Ground state and vortex states of bosons in an
anisotropic trap: A variational approach} 
\author{Manoranjan P. Singh and A. L. Satheesha\cite{satish} } 
\address{Laser Programme, Centre for Advanced Technology, Indore 452013,
India}
\date{\today}
\maketitle 
\draft 
\begin{abstract}
We propose a simple variational form of the wave function to
describe the ground state and vortex states of a system of
weakly interacting Bose gas in an anisotropic trap. The proposed wave
function is valid for a wide range of the particle numbers in the trap. 
It also works well  in
the case of attractive interaction between the atoms.  Further, it
provides an easy and fast method to calculate the physical
quantities of interest. The results compare very well with those
obtained by purely numerical techniques. Using our wave function we
have been able to verify, for the first time, the predicted
behaviour of aspect ratio.
\end{abstract} 


\section{Introduction}
Observation of Bose-Einstein condensation in cooled and trapped
dilute gases of alkali atoms \cite{exptrb,exptli,exptna} and spin polarized
atomic hydrogen\cite{expth} has generated a renewed
theoretical interest in understanding such systems. In a
meanfield approach, which is valid in the limit $\rho a^{3} <<
1$, where $\rho$ is density of atoms and $a$ is the $s$-wave
scattering length, ground state and vortex states of these
systems can be described by Gross-Pitaevskii (GP) equation\cite{grossp}.
Various numerical procedures\cite{sym1,sym2,asy,dal,li} and approximate
analytical methods\cite{baym,var,moroz,fetter,effpot} have been used to solve
the GP equation. Among these variational scheme proposed by Baym and Pethick
\cite{baym} to explain the experimental observations of Ref.\cite{exptrb} is
particularly appealing. In this approach the trial  wave function was taken
to be of the form of ground state of the trap potential (modeled
by an anisotropic harmonic oscillator potential).  Thus the wave function is represented
by a three dimensional
Gaussian with axial and transverse frequencies as variational
parameters.  This form of wave function, however, is
valid only when the number of atoms in the trap is very small. As
the number increases, the repulsive interaction between the atoms
tends to expand the condensate and flatten the density profile
in the  central region of the trap where the density is maximum.  
Of these two effects, only expansion of the
condensate can be described adequately by the Gaussian trial wave function.
On the other hand, the Thomas-Fermi
approximation \cite{dal,baym} provides a wave function which is valid when the
number of atoms is very large which is the case with  the recent experiments 
\cite{recent}. It is of interest, however, to have a form of wave 
function which  apart from showing these limiting behaviours, is also valid
 in the intermediate region. In a variational approach, Fetter \cite{fetter} has 
proposed such a wave function for the bosons in an isotropic trap.
 
In this paper we propose a simple form of wave function for the
ground state of bosons confined in an anisotropic trap. The wave function is 
valid  for a very wide range of  particle numbers.
When the number is small it tends to mimic a Gaussian,
and in the opposite limit it resembles the Thomas-Fermi wave
function. However for large number of atoms the wave function differs from the
Thomas-Fermi wave function in the surface region, a desirable feature as noted
in Ref.\cite{dal,dal1}.  The trial wave function
has got an additional parameter comapared to the ones used in
Ref.\cite{baym}. This parameter takes care of the flattening of
the density in the central region of the condensate. Thus, it provides
a better lower bound for the ground state energy than the Gaussian
trial wave function. We also compare the results obtained by our
trial wave function with those obtained by other numerical
procedures such as the minimization of energy functional by
steepest descent method \cite{dal} and the integration of the
nonlinear Schrodinger equation \cite{asy,li}. These comparisons show the
form chosen by us to be highly accurate for obtaining a host of  physical
qauantities of interest. 
In addition
to providing acurate results for a very wide range of the particle numbers,
the method is also very fast from the computational point of
view.  Further, using  this wave function, the physical observables can be 
expressed analytically in
terms of three variational parameters which are  obtained
by minimizing the GP energy functional. Since we have a simple
analytical form for the energy in terms of three variational
parameters the procedure of minimization is very simple. 
The novel achievement of this
method lies in verfication of the predicted behaviour of the
aspect ratio which is very important quantity from the experimental
point of view. It could not be ascertained before because
convergence of the aspect ratio to the highly repulsive limit is very 
slow \cite{dal}.
Since our method can handle even very large number of atoms in the
trap without any difficulty we could verify the behaviour of the
aspect ratio.

Based on physical reasoning we generalise the variational form
to descibe the vortex states also. As is the case with the ground 
state, we find good agreement with the existing results, 
with considerably less computational effort.

Condensation has also been observed in $^7$Li\cite{exptli} where
 interatomic interaction is attractive which
is characterized by the negative s-wave scattering length. As
the number of atoms in the trap inceases the condensate shrinks
and nonuniformity in the central region increases. After a
critical number of atoms in the trap, the condensate collapses.
This situation is also very well descibed by our trial wave
function. The parameter which accounts for flattening of the
density profile in the case of repulsive interactions also takes care
of the increase in density gradient in an effective way . The
critical number of atoms for the case of $^7$Li is in close
agreement with that reported in Ref.\cite{li}.

The paper is organized as follows. Section II contains
the description of the variational scheme employed in the paper.
It also contains the analytic expressions for the observables of
interest. Results obtained from the variational procedure and
their comparison with the existing ones are presented in Section
III. Section IV contains summary and  concluding remarks.

\section{Variational Method}
\subsection{Ground state}
Bose Einstein condensation in  experiments with cooled and
trapped atoms can be described within  the framework of the GP
theory.  Validity of such a description has been analysed by
Stenholm\cite{valid}. In a situation where the trap can be
modeled by an anisotropic harmonic oscillator potential characterized by
the two angular frequencies $\omega^0_{\bot}$ and $\omega^0_z$, and the
interatomic ineractions can be replaced by an effective
pseudo-potential involving $s$-wave scattering length $a$, the ground
state energy for condensed bosons of mass $m$ is given by the
 Gross-Pitaevskii functional\cite{grossp}
\begin{equation}
\frac{E_{1}[\psi]}{N}=\int{d{\bf r}_{1}\frac{1}{2}
\left[|\nabla_{1} \psi_{1}({\bf r_{1}})|^2
+ \left( x_{1}^2 + y_{1}^2 +\lambda_{0}^2 z_{1}^2
\right)|\psi({\bf  r}_{1})|^2
+\frac{u_{1}}{2} |\psi_{1}({\bf r}_{1})|^4\right]}.
\label{eq6}
\end{equation}
Here we have used the length scale $a_{\bot}=\sqrt{{\hbar}/{m \omega_{\bot}^0}}$
 and the energy scale $\hbar \omega^{0}_{\bot}$ provided by the trap potential 
to express Eq.(\ref{eq6}) in terms of the dimensionless variables. 
$\psi_{1}({\bf r}_{1})$ is the condensate wave function 
which satisfies  the normalization condition
\begin{equation}
\int {d{\bf r}_{1} |\psi_{1}({\bf r}_{1})|^2} = 1.
\label{eq8}
\end{equation}
$\lambda_{0}={\omega_{z}^0}/{\omega_{\bot}^0}$ is the anisotropy parameter 
of the trap, $N$ is the total number of atoms in the condensate and 
$u_1 = {8 \pi a N}/{a_{\bot}}$.   The exact form of the wave function can be 
determined by minimizing the
energy functional in Eq.(\ref{eq6}) with the normalization constraint
of Eq.(\ref{eq8}). Such a minimization results in  the
nonlinear Schrodinger equation
\begin{equation}
\left[- \nabla_{1}^2 
+ \left( x_{1}^2 + y_{1}^2 + \lambda^{2}_{0} z_{1}^2 \right)
+u_{1} |\psi_{1}({\bf r}_{1})|^2\right]\psi_{1}({\bf r}_{1}) = 2
\mu_{1} \psi_{1}({\bf r}_{1}),
\label{eq9}
\end{equation}
where, $\mu_1$ is the chemical potential.
It is not possible to find exact analytic solution to
Eq.(\ref{eq9}).  Consequently various numerical techniques have
been developed to study the ground state property of such systems within the
framework of the GP theory. These techniques involve either the direct
numerical minimization of  Eq.(\ref{eq6})  with the constraint of 
Eq.(\ref{eq8}) \cite{dal} or numerical integration of Eq.(\ref{eq9}) or its
 time dependent version\cite{sym1,sym2,asy,li}. Another approach is to use the
variational method which has been extensively used in 
different branches of physics. The main advantage of this method is that with
 a suitable guess
for the form of the wave function it is possible to save a lot
of computational effort and time. In addition, it may also
provide physical insights which generally get obscured in the
complicated computational procedures. The first study of this
kind was done by Baym and Pethick \cite{baym} in  light of
the experimental observations in $^{87}$Rb \cite{exptrb}. They took
the trial wave function for the ground state as
\begin{equation}
\psi({\bf r})=N^{1/2} \omega_{\bot}^{1/2} \omega_{z}^{1/4} \left(
\frac{m}{ \pi \hbar}\right)^{3/4}
 e^{-m(\omega_{\bot} r_{\bot}^2 +\omega_{z} z^2)/{2 \hbar}}
\label{eq10}
\end{equation}
with effective frequencies, $\omega_{\bot}$ and $\omega_z$,
treated as variational parameters. However, the wave function above  brings
 out only the qualitative features of
the condensate {\em e.g.} expansion of the condensate in
different directions, shifts in the angular frequencies and the
scaling behaviour of energy with the number of atoms in the trap. Further,
 this form of the wave function is valid
only for  small number of atoms in the trap (see Fig.2 below).
 We now propose a variational
form of the wave function and demonstrate its applicability and
utility in providing accurate description of the condensate for a wide
range of the particle numbers.
 The form of the trial wave function we choose is
\begin{equation}
\psi_{1}({\bf r}_{1})= \sqrt{\frac{p}{2 \pi \Gamma(\frac{3}{2 p})}}
\lambda^{\frac{1}{4}} \left(\frac{\omega_{\bot}}{\omega_{\bot}^0}\right)
^{\frac{3}{4}} e^{-\frac{1}{2}
\left(\omega_{\bot}/\omega_{\bot}^0\right)^p
\left( r_{1 \bot}^2 +\lambda z_1^2\right)^p} ,
\label{eq11}
\end{equation}
 where, $\lambda,
\omega_{\bot}$ and $p$
are the variational parameters which are obtained by minimizing
the energy $E_1$ in Eq.(\ref{eq6}) with respect to these
parameters. It is easily  verified that  the wave function satisfies the
normalization condition of Eq.(\ref{eq8}).  The
 expression of the ground state energy $E_1$ in terms of
 $\lambda$, $\omega_{\bot}$ and $p$ is
\begin{eqnarray}
E_{1}&=&\frac{1}{12}\frac{\omega_{\bot}}{\omega^0_{\bot}}
\left(1+\frac{\lambda}{2}\right)
\frac{\Gamma\left( \frac{1}{2p}\right)}{\Gamma\left(\frac{3}{2p}\right)}
(1+2p) +\frac{1}{3}\frac{\omega^0_{\bot}}{\omega_{\bot}}
\left(1+\frac{\lambda_0^2}{2\lambda}\right)
\frac{\Gamma\left(
\frac{5}{2p}\right)}{\Gamma\left(\frac{3}{2p}\right)}\nonumber \\ 
& &+N\frac{a}{a_{\bot}}
\left(\frac{\omega_{\bot}}{\omega^0_{\bot}}\right)^{\frac{3}{2}}
\frac{\sqrt{\lambda}
p}{\Gamma\left(\frac{3}{2p}\right)}
\left(\frac{1}{2}\right)^{\frac{3}{2p}} .
\label{eq12}
\end{eqnarray}
For a particular value of $N$ the  parameters $\omega_{\bot}$, $\lambda $
 and $p$ are obtained
 by minimizing the energy above using standard numerical routines.
 We have used {\em Mathematica}\cite{math} for this and
it takes a few seconds of the real time on 166 MHz Pentium-1 computer to get 
 the answer. Next we discuss how different 
 physical observables can be obtained in terms of the
 parameters of the wave function. Aspect ratio which characterizes the
anisotropy of the velocity distribution of the condensate is
defined as $\sqrt{<p_z^2>/<p_x^2>}$. This can be easily shown to be 
\begin{equation}
\sqrt{\frac{<p_z^2>}{<p_x^2>}}=
\sqrt{\frac{<x_1^2>}{<z_1^2>}}= \sqrt{\lambda} .
\label{eq13}
\end{equation}
Width of the condensate in the transervse direction is given by
\begin{equation}
<x_1^2>=\frac{\omega_{\bot}^0\Gamma\left(\frac{5}{2 p}\right)}
{3 \omega_{\bot}\Gamma\left(\frac{3}{2 p}\right)}
\label{eq14}
\end{equation}
and the width of the the momentum distribution in this direction
is given by
\begin{equation}
<p_x^2>=\frac{N \hbar m \omega_{\bot}\Gamma\left(\frac{1}{2 p}\right)
(1+2 p)}
{12 \Gamma\left(\frac{3}{2 p}\right)} .
\label{eq15}
\end{equation}
The peak density of the condensate is $N \psi_1^2(0)/a_{\bot}^3$. Life time
of the condensate is related to the density distribution. The loss rate
due to the two body loss rate\cite{loss2} and the three body loss rate
\cite{loss3} is given by
\begin{eqnarray}
R(N)&=&\alpha \int d{\bf r} |\psi({\bf r})|^4 
+L \int d{\bf r} |\psi({\bf r})|^6 \nonumber \\
&=& \frac{\alpha N^2 \sqrt{\lambda} (\omega_{\bot}/\omega_{\bot}^0)^{3/2} p}
{2 \pi 2^{3/2p} a_{\bot}^3 \Gamma(3/2p)} 
+ \frac{L N^3 \lambda (\omega_{\bot}/\omega_{\bot}^0)^3 p^2}
{4 \pi^2 3^{3/2p} a_{\bot}^6 \Gamma^2(3/2p)} ,
\label{lossrate}
\end{eqnarray}
where, $\alpha$ is the two-body dipolar relaxation loss rate coefficient
and $L$ is the three-body recombination loss rate coefficient.
\subsection{Vortex States}
We consider here the states having a vortex line along the $z$
axis. Wave function of such a state can be written as
\begin{equation}
\Psi ({\bf r})= 
\psi ({\bf r})  e^{\imath \kappa \phi}
\label{eq16}
\end{equation}
where $\kappa$ is an integer denoting the quantum of
circulation. Subsituting the complex wave function $\Psi$ in
place of $\psi$ in Eq.(\ref{eq6}) we get Gross-Pitaevskii
functional for the vortex states in terms of the scaled variables
\begin{equation}
\frac{E_{1}[\psi]}{N}=\int{d{\bf r}_{1} \frac{1}{2} \left[|\nabla_{1} 
\psi_{1}({\bf r_{1}})|^2
+ \left( \kappa^2 r_{1 \bot}^{-2} + r_{1 \bot}^2 +\lambda_{0}^2
z_{1}^2
\right)|\psi({\bf  r}_{1})|^2
+\frac{u_{1}}{2} |\psi_{1}({\bf r}_{1})|^4\right]}.
\label{eq17}
\end{equation}
The corressponding nonlinear Schrodinger equation is
\begin{equation}
\left[- \nabla_{1}^2 
+ \kappa^2 r_{1 \bot}^{-2} + r_{1 \bot}^2 +\lambda_{0}^2 z_{1}^2
+u_{1} |\psi_{1}({\bf r}_{1})|^2\right]\psi_{1}({\bf r}_{1}) = 2
\mu_{1} \psi_{1}({\bf r}_{1}).
\label{eq18}
\end{equation}
We assume the trial wave function of the form
\begin{equation}
\psi_{1}({\bf r}_{1})  = A r_{1\bot}^q
e^{-\frac{1}{2}
\left(\omega_{\bot}/\omega_{\bot}^0\right)^p
\left( r_{1 \bot}^2 +\lambda z_1^2\right)^p}
\label{eq19}
\end{equation}
where $q$ is an  additional variational parameter.  This
particular form of the wave function is motivated by the
following considerations.
\begin{enumerate}
\item{Presence of the centrifugal term $\kappa^2/r_{1\bot}^2$ forces the
wave function to vanish on the $z$ axis.}
\item{It has been shown that for a weakly interacting Bose gas \cite{pathria}
the wave function corresponding to $k$th quantum circulation
behaves as
\begin{equation}
\psi \sim r_{1\bot}^k
\label{eq20}
\end{equation}
near the $z$ axis.}
\end{enumerate}
Proportionality factor in Eq.(\ref{eq19}) is determined by the
normalization condition (Eq.\ref{eq8})
\begin{equation}
A^2=\frac{\sqrt{\lambda} p \Gamma\left(\frac{3}{2} +q \right)}
{\pi^{3/2} \Gamma(1+q) \Gamma\left(\frac{3 +2q}{2p}\right)}
 \left(\frac{\omega_{\bot}}{\omega_{\bot}^0}\right)^{(3+2q)/2}.
\label{vornorm}
\end{equation}
For a vortex line descibed by  the wave function in Eq.(\ref{eq19})
the density peaks at
\begin{equation}
 r_{1\bot}= \sqrt{\frac{\omega_{\bot}^0}{\omega_{\bot}}} \left(\frac{q}{p}\right)^
{\frac{1}{2p}}, 
\label{vorpeak}
\end{equation} 
and the peak density is  given by
\begin{equation}
\rho_{max}=\frac{N}{a_{\bot}^3} A^2
 \left(\frac{\omega_{\bot}^0}{\omega_{\bot}}\right)^q \left(\frac{q}{p}\right)^
{\frac{q}{p}} e^{-q/p} .
\label{vordens}
\end{equation} 
 It is also sraightforward
to get the analytic expression for $E_1$ in terms of the
variational parameters $\omega$, $\lambda$, $p$ and $q$ which,
in turn, are obtained by minimization of $E_1$. The kinetic energy is
given by  
\begin{equation}
(E_1/N)_{kin}=\frac{\omega_{\bot} (1+2q)
 \left[(1+2p)(1+\lambda/2) +q(2p+2q+\lambda)\right]
  \Gamma\left(\frac{1 +2q}{2p}\right)}
{4 (3+2q) \omega_{\bot}^0
  \Gamma\left(\frac{3 +2q}{2p}\right)}.
\label{vorkin}
\end{equation}
Energy corresponding to the rotational motion is given by
\begin{equation}
(E_1/N)_{rot}=
\frac{\kappa^2 \omega_{\bot} (1+2q)
  \Gamma\left(\frac{1 +2q}{2p}\right)}
{4 \omega_{\bot}^0 q
  \Gamma\left(\frac{3 +2q}{2p}\right)}.
\label{vorrot}
\end{equation}
The oscillator energy is given by
\begin{equation}
(E_1/N)_{HO}=
\frac{\omega_{\bot}^0 \lambda \left(1+q+\frac{\lambda_0^2}{2 \lambda}\right)
  \Gamma\left(\frac{5 +2q}{2p}\right)}
{\omega_{\bot} \left(3+2q\right)
  \Gamma\left(\frac{3 +2q}{2p}\right)} .
\label{vorho}
\end{equation}
The interaction energy is given by
\begin{equation}
(E_1/N)_{pot}=
 2 \left(\frac{\omega_{\bot}}{\omega_{\bot}^0}\right)^{3/2}
\frac{a\sqrt{\lambda} N p (1+2q)^2 \Gamma(2q) 
  \Gamma^2\left(\frac{1}{2}+q\right)
  \Gamma\left(\frac{3 +4q}{2p}\right)} 
{2^{(3+4q)/2p} \pi^{1/2} a_{\bot} q (1+4q) \Gamma^2(q)
  \Gamma\left(\frac{1}{2}+2q\right)
  \Gamma^2\left(\frac{3 +2q}{2p}\right)}.
\label{vorpot}
\end{equation}
It is easy to verify that ground state is obtained by setting $\kappa$ and
$q$ equal to zero. Once we have the  energy of the states with and without
vortices we can calculate the critical angular velocity for the formation
of the vortex state. In the unit of $\omega_{\bot}^0$ it is given by \cite
{dal}
\begin{equation}
\Omega_c=\kappa^{-1} \left[(E_1/N)_{\kappa}-(E_1/N)_0\right].
\label{omegac}
\end{equation} 

 To demonstrate the applicability of this method we
have performed calculations for $^{87}$Rb and $^{7}$Li. The $s$
wave scattering length is positive for $^{87}$Rb. It is negative
for $^{7}$Li. Consequently the interatomic interaction is
repulsive in the former and attractive in the latter. We now
present the results and their comparison with the existing
calculations.

\section{Results}
\subsection{Positive scattering length: $^{87}$Rb}
In this section we report calculations on $^{87}$Rb. We perform our calculations by employing
the experimental numbers for the asymmetry parameter of the trap, the axial
 frequency and the $s$ wave scattering length
 corresponding to the
experimental situation of Ref.\cite{exptrb} and the subsequent
theoretical calculations \cite{asy,dal,baym}. Accordingly,  $\lambda^0=
\omega^0_z/\omega^0_{\bot} =\sqrt{8}$;  $\omega^0_z/{2 \pi}
$ is  220 Hz;  $a$ is
100$a_0$, where $a_0$ is the Bohr radius. The corresponding
characteristic length is $a_{
\bot} = 1.222 \times 10^{-4}$ cm and the ratio $a/a_{\bot}$ is $4.33\times 
10^{-4}$.

First, we obtain the energy $E_1$ (Eq.\ref{eq12}) by minimizing it 
with respect to the variational parameters $\omega_{\bot}$,
$\lambda$ and $p$ for various values of the particle number $N$.
We present the results in Fig.1. It is evident that the results obtained by us
 are in close agreement with
the results in  Ref.\cite{dal}(see also Table 1 below). As pointed out above, 
 these agree well with the
results obtained by using the Gaussian trial wave function when $N$ is
small and with those obtained by using 
the Thomas-Fermi approximation\cite{dal,baym} when $N$ is very
large. These comparisons clearly establish the validity of our wave function
for a very wide range of the particle numbers.

Next, we compare the proposed wave function with the 
Gaussian trial wave function  of Eq.(\ref{eq10})
and also the one given by the Thomas-Fermi
approximation in Fig.2. It is clear
that when the number of atoms in the trap is small the proposed
wavefunction has more resemblence with the Gaussian wavefunction
(Fig.2a). As $N$ is increased the wave function tends to flatten
in the central region and the resulting form is a mixture of the
two limiting forms i.e. the Gaussian and the Thomas-Fermi wave
function.  In the central region it is close to  the
latter while it resembles  the former away from the trap
centre (Fig.2b).  When $N$  is very large the
resemblence is more with the Thomas-Fermi wave function (Fig.2c). However
we note that the wave function 
 vanishes smoothly far away from the centre of the trap. As mentioned
above this a desirable feature which is crucial for the calculation of
some relevent physical observables {\em e.g.} the  aspect ratio \cite{dal}. It is clear that
our wave  function  not only provides a better lower bound
for the energy but also shows the correct and the desirable limiting
behaviour. 

Results of calculation of various quantities {\em e.g.} chemical potential,
total energy , kinetic energy, potential energy, interaction
energy, average size of the condensate in the tranverse
direction and in the axial direction have been presented in Table
1. The close agreement with the results of Ref.\cite{dal} is evident.
 We have also looked at  the variation of the peak density and the
total loss rate of the atoms with the particle number $N$. We find them
to be in good agreement with the result of  Ref.\cite{asy}. 

We now present the behaviour of the aspect ratio which is a very important
quantity from  the experimental point of view . As mentioned in
Ref. \cite{dal,baym} it is equal to $\sqrt{\lambda_0}$ in the
non-interacting limit and tends to $\lambda_0$ in the highly repulsive
limit, which is the case when $N$ is very large. However the
convergence to the repulsive limit is very slow \cite{dal,baym} and
therefore this behaviour has not been seen explicitly so far. On the other hand,
with a variational wave function, calculations can be performed for any $N$
with equal ease.
 Consequently we have been able to verify the predicted
behaviour of the aspect ratio. It is seen from Fig.3 that it
requires calculations up to $N \sim 10^6$ to see the aforesaid
behaviour.

In Fig.4 we show the  wave function of the vortex state corresponding to
 $\kappa=1$ for $N=5000$. The atoms are pushed away from the $z$ axis.
Peak density is 7.155$\times 10^{13}$ which occurs at $r_{1\bot}=1.611$.
The position of the peak moves further away from the $z$ axis
as $N$ is increased.   For $N=10000$ the peak  occurs at
 $r_{1\bot}=1.844$ while it is 
 at $r_{1\bot}=2.161$ for $N=20000$. Value of  the peak density 
remains much the same for $N=10000$ and 20000( 9.524$\times 10^{13}$ and
 12.26$\times 10^{13}$, respectively). We also compare the energy of the 
vortex state with different values of $\kappa$ with that obtained in the 
Thomas-Fermi approximation \cite{sinha}. For calculation in the Thomas-Fermi 
approximation we have neglected the change in the chemical potential due to 
the presence of a vortex line. Results are  presented in Table 2. The 
Thomas-Fermi approximation yields lower value of the energy which is expected 
as it ignores the  kinetic energy contribution from the core and the surface 
regions of the vortex state. This approximation is valid only for very large
 $N$. We do find  
the results of the two calculations to be in close agreement for large $N$. For
$N=10^6$ the two results differ only by $\approx 1\%$. We have also calculated
 the critical angular velocity  for the vortex state  with $\kappa=1$ for 
various values of the particle number $N$. We find that it decreases rapidly in the begining (up to $N \sim
2000$). Thereafter it varies slowly.  The result shows good quantitative
agreement with those  in the Ref.\cite{asy,dal}. For example for
 $N=2000$ the critical angular velocity
is 52$\%$ of the noninteracting value given by  the transverse angular
frequency $\omega_{\bot}^0$ of the trap,  in comparison with
 49.33$\%$ in Ref.\cite{asy}.
For $N>5000$ it is less than 43$\%$ of the noninteracting value, which compares
well with the figure of 40$\%$  in   Ref.\cite{dal}. The critical angular
velocity increases with $\kappa$. For $N=10000$ we find $\Omega_c/2\pi$ equal
to 30.57, 38.5, and 45.42 Hz for $\kappa =$1, 2, and 3 respectively. These
figures are 26, 35, and 41 Hz respectively in Ref.\cite{dal}.

\subsection{Negative scattering length: $^{7}$Li}
In this section we report calculations on $^{7}$Li. These atoms interact
via an attractive interaction and consequently the scattering length in this
case is  a negative quantity.
Numerical values of the parameters used in the calculations correspond
to the experimental situation of Ref.\cite{exptli} and the
subsequent theoretical calculations \cite{dal,li}. Accordingly,
the asymmetry parameter of the trap is $\lambda^0=
\omega^0_z/\omega^0_{\bot} =0.72$. The axial frequency $\omega^0_z/{2 \pi}
$ is taken to be 117 Hz. The $s$-wave scattering length $a$ is
$-27a_0$
. The corresponding characteristic length is $a_{
\bot} = 2.972 \times 10^{-4}$ cm and the ratio $a/a_{\bot}$ is $- 4.33\times 
10^{-4}$.
 
We find the value of the critcal number $N_c=1270$ beyond which
the ground state collapses because of the attractive
interaction. This is in good agreement with the figure of $N_c \sim 1300$
in Ref.\cite{li} and the experimental observation (see the second paper of 
Ref.\cite{exptli}) that $N \le 1300$. Wave functions for $N=$ 500 and 1270 are shown in 
Fig.5. For $N =$  500 there is hardly any difference between the proposed
wave function and the Gaussian trial wave function. However, the difference
is rather significant for $N = $ 1270 as is evident from the figure. We plot 
the aspect ratio for various
values of $N \le N_c$ in Fig.6. At $N \sim N_c$ the aspect
ratio tends to 1. Since for a wave function of the form given by
Eq.(\ref{eq11})  the aspect ratio also gives ratio of spatial
widths in the transverse and the axial directions, the
condensate tends to be isotropic for $N \sim N_c$.  This becomes
further evident in Table 3 where we have listed the results for various
quantities of interest. As reported in Ref.\cite{dal} the variation in the
various quantities is smooth from $N$ = 1 to 1000. However, we can also note the sharp  variation as we reach 
 the critical number.  This
behaviour is consistent with that reported in Ref.\cite{li}. We have also 
calculated the peak density and loss rate for $N \le N_c$. Once
again we find a sharp increase near $N \sim N_c$. These results
 also match well with those of Ref.\cite{li}.

It is  possible to have very large number of particles ($N >> 1300$) in
the vortex  states even when the interatomic interaction is attractive. We 
have considered the vortex states with $\kappa=$ 1, 2 and 3. The particle number is
3500, 6000 and 8000 respectively. Peak densities for these states are $1.266
\times 10^{13}$, $2.239 \times 10^{13}$ and $2.744 \times 10^{13}$ which
occur at $r_{1\bot} =$ 0.922, 1.257 and 1.571 respectively. Although the
particle number is quite different in the three cases, the peak densities are
not  very different. Also they remain  less than the peak density $3.984 
 \times 10^{13}$ which corressponds to $\kappa=0$ and $N=1270$. These
 observations are consistent with those in Ref.\cite{dal}. Stability of the
vortex states for attractive interaction can be physically explained as the
interplay between the restoring force and and the centrifugal force. The restoring
force tries to attract the particles to the centre while the centrifugal
force tries to push them out. The net effect is that  the peak density does not
change much even when there is siginificant variation in the particle number.
Since the interparticle interaction depends on the density, for low dnesities
it does not cause the collapse of the condensate.

In case of attractive interaction it takes more energy to create a vortex
state than that requiread in the noninteracting case. Consequenty the critical
angular velocity is greater than  unity. For $\kappa=1$ and $N=1000$ we find 
 $\Omega_c = 1.119$ which compares very well with $\Omega_c=1.2$ reported
in Ref.\cite{dal}.

\section{Conclusion}

We have proposed a variational scheme to describe  the ground
state and vortex states of weakly interacting atomic gases
confined by harmonic traps within the framework
of the meanfield theory of Gross and Pitaevskii . It is based on a judicious choice of the form of
trial  wave function for the ground state
 which has a simple functional form and at the same time
 is  valid for a  wide range of the particle numbers. When the number
 is small it tends towards a Gaussian and in the opposite limit it resembles the
Thomas-Fermi wave function. However, for large $N$  it provides a better 
description of 
 the surface region than the Thomas-Fermi wave function. In the intermediate regime
it combines the feature of  both in an effective way. In the central region
of the trap, where density is high, it matches with the Thomas-Fermi wave
 function. Away from the centre of the trap, where density is low, it matches
the Gaussian trial wave function. We easily generalize the wave function for the 
vortex states.  We have demonstrated the
applicability of our method by performing calculations of  various physical quantities
 for the experimental
situations of Ref.\cite{exptrb} and Ref.\cite{exptli}. We find our results 
to be in
good agreement with the existing results. The method is semi-analytic and 
consequently
 computationally easy to implement. As our method poses no additional
problems even  for very large particle numbers we have been able to verify,
 for the first time, the predicted behaviour of the aspect ratio. The formalism
 is quite general involving only the scaled $s$-wave scattering length
and the asymmetry parameter of the trap. We therefore believe that it will
be useful in analysing a variety of experiments. In addition it may
serve as a very good starting point for the theories\cite{kagan} where quantum
fluctuations play an important role. Generalization of our method to the 
time dependent case \cite{var} is straightforward and it will be reported in 
a future publication.

\section{acknowledgment}
We thank Dr. S. C. Mehendale and Dr. M. K. Harbola  for helpful discussions and critical reading
of the manuscript.

\newpage
\begin{table}
\caption[test]{Results for the ground state of $^{87}$Rb atoms confined in an
anisotropic harmonic trap with $\lambda_0=\sqrt{8}$ and $\omega^0_{\bot}/2\pi= 220/\lambda_0$ Hz. Chemical potential and energy are in units of
$\hbar \omega^0_{\bot}$ and length is in  units $a_{\bot}$. Numbers in the
brackets correspond to the results of Ref.\protect\cite{dal}.}
\begin{tabular}{lccccccc}
$N$ & $\mu_{1}$ & $(E_1/N)$ & $(E_{1}/N)_{kin}$ &
$(E_{1}/N)_{HO}$&$(E_{1}/N)_{pot}$
&$\sqrt{<x_1^2>}$&$\sqrt{<z_1^2>}$ \\ \hline
1&2.42&2.417&1.205&1.209&0.003&0.708&0.421 \\
&(2.414)&(2.414)&(1.207)&(1.207)&(0.000)&(0.707)&(0.42) \\
100&2.88&2.663&1.06&1.39&0.217&0.788&0.437 \\
&(2.88)&(2.66)&(1.06)&(1.39)&(0.21)&(0.79)&(0.44) \\ 
200&3.219&2.859&0.98&1.52&0.36&0.845&0.45 \\
&(3.21)&(2.86)&(0.98)&(1.52)&(0.36)&(0.85)&(0.45) \\ 
500&3.95&3.309&0.854&1.815&0.641&0.959&0.473 \\
&(3.94)&(3.30)&(0.86)&(1.81)&(0.63)&(0.96)&(0.47) \\ 
1000&4.787&3.851&0.755&2.16&0.936&1.078&0.499 \\
&(4.77)&(3.84)&(0.76)&(2.15)&(0.93)&(1.08)&(0.5) \\ 
2000&5.951&4.628&0.66&2.645&1.323&1.227&0.534 \\
&(5.93)&(4.61)&(0.66)&(2.64)&(1.32)&(1.23)&(0.53) \\
5000&8.164&6.142&0.543&3.577&2.022&1.469&0.596 \\
&(8.14)&(6.12)&(0.54)&(3.57)&(2.02)&(1.47)&(0.59) \\ 
10000&10.527&7.783&0.461&4.577&2.744&1.689&0.657 \\
&(10.5)&(7.76)&(0.45)&(4.57)&(2.74)&(1.69)&(0.65) \\ 
15000&12.264&8.999&0.416&5.317&3.266&1.833&0.699 \\
&(12.2)&(8.98)&(0.41)&(5.31)&(3.26)&(1.84)&(0.7) \\ 
20000&13.689&9.998&0.385&5.922&3.691&1.944&0.732 \\
&(13.7)&(9.98)&(0.38)&(5.91)&(3.68)&(1.94)&(0.73) \\
\end{tabular}
\end{table}
\newpage
\begin{table}
\caption[test]{Results for the energy of vortex states with $\kappa=1, 2$ and
3 of $^{87}$Rb atoms confined in an anisotropic harmonic trap with 
$\lambda_0=\sqrt{8}$ and $\omega^0_{\bot}/2\pi= 220/\lambda_0$ Hz. 
Energy is in units of
$\hbar \omega^0_{\bot}$. Numbers in the brackets correspond to the results 
obtained in the Thomas-Fermi limit.}
\begin{tabular}{lccc}
$N$ & $(E_1/N)_{\kappa=1}$ & $(E_{1}/N)_{\kappa=2}$ &
$(E_{1}/N)_{\kappa=3}$ \\ \hline
10$^3$& 4.455& 5.274& 6.164 \\
& (3.385)& (4.057)& (4.402) \\
$5 \times 10^3$& 6.579& 7.237& 7.993 \\
& (5.845)& (6.513)& (7.219) \\
$10^4$& 8.176& 8.773& 9.469 \\
& (7.537)& (8.146)& (8.853) \\
$5 \times 10^4$& 14.455& 14.94& 15.511 \\
& (13.951)& (14.397)& (14.977) \\
$10^5$& 18.78& 19.227& 19.771 \\
&(18.306) & (18.684)& (19.192) \\
$5 \times 10^5$& 35.097& 35.474& 36.072 \\
& (34.627)& (34.876)& (35.227) \\
$10^6$& 46.134& 46.485& 46.888 \\
& (45.637)& (45.842)& (46.134) \\
$5 \times 10^6$& 87.416& 87.72& 88.061 \\
& (86.764)& (86.892)& (87.078) \\
$10^7$& 115.229& 115.518& 115.837 \\
& (114.459)& (114.562)& (114.714) \\
\end{tabular}
\end{table}
\newpage
\begin{table}
\caption[test]{Results for the ground state of $^{7}$Li atoms confined in an
anisotropic harmonic trap with $\lambda_0=0.72$ and $\omega^0_{\bot}/2\pi= 163$ Hz. Chemical potential and energy are in units of
$\hbar \omega^0_{\bot}$ and length is in  units $a_{\bot}$. Numbers in the
brackets correspond to the results of Ref.\protect\cite{dal}.}
\begin{tabular}{lccccccc}
$N$& $\mu_{1}$ & $(E_1/N)$& $(E_{1}/N)_{kin}$&
$(E_{1}/N)_{HO}$&$(E_{1}/N)_{pot}$
&$\sqrt{<x_1^2>}$&$\sqrt{<z_1^2>}$ \\ \hline 1& 1.36& 1.36&
0.68& 0.68& 0& 0.707& 0.833 \\ & & (1.36)& & & & & \\ 
100& 1.327& 1.344& 0.693& 0.670& -0.017& 0.701& 0.824 \\ 200&
1.291& 1.326& 0.707& 0.654& -0.035& 0.695& 0.813 \\ 300& 1.254&
1.309& 0.722& 0.641& -0.054& 0.688& 0.803 \\ 400& 1.214& 1.29&
0.74& 0.626& -0.076& 0.681& 0.791 \\ 500& 1.173& 1.271& 0.758&
0.611& -0.098& 0.672& 0.786 \\ 600& 1.125& 1.25& 0.782& 0.594&
-0.125& 0.665& 0.765 \\ 700& 1.074& 1.229& 0.808& 0.576& -0.155&
0.656& 0.75 \\ 800& 1.017& 1.206& 0.839& 0.556& -0.189& 0.645&
0.734 \\ 900& 0.952& 1.182& 0.878& 0.533& -0.23& 0.633& 0.715 \\
1000& 0.874& 1.155& 0.928& 0.507& -0.28& 0.619& 0.693 \\ & &
(1.15)& & & &(0.62)& (0.69) \\  1100& 0.776& 1.125& 0.999&
0.475& -0.349& 0.60& 0.665 \\ 1200& 0.625& 1.09& 1.121& 0.43&
-0.461& 0.573& 0.625 \\ 1270& 0.346& 1.06& 1.42& 0.352& -0.713&
0.521& 0.554 \\
\end{tabular}
\end{table}
\newpage
\begin{center} \bf Figure Captions \end{center}
\begin{itemize}
\item[FIG. 1.]Ground state energy per atom for $^{87}$Rb as a
function of $N$. Energy is in the units of $\hbar
\omega_{\bot}^0$.  The solid line is the result of our
variational calculation. The dashed line is the result obtained
by using the Gaussian trial wave function of
Ref.\cite{baym}, while the dotted line is the result 
obtained by using the Thomas-Fermi approximation. The filled circles
are the results of Ref.\cite{dal}.

\item[FIG. 2.]Ground state wave function for $^{87}$Rb along the $x$ axis for
 different values of $N$. The  solid line is the result of
 our variational calculation.
The dotted  line is the  result obtained by using the Gaussian trial
wave function of Ref.\cite{baym}, while the dashed line is the result
obtained by using the Thomas-Fermi approximation.(a) $N=100$. (b)$N=1000$.
 (c)$N=50000$.

\item[FIG. 3.]Aspect ratio in $^{87}$Rb as a function of $N$. The lower and upper 
 horizontal lines correspond to $\sqrt{\lambda_0}$ and $\lambda_0$ respectively.

\item[FIG. 4.]Vortex-state wave function of 5000 $^{87}$Rb atoms along the
$x$ axis for $\kappa=1$. 

\item[FIG. 5.]Ground state wave function for $^{7}$Li along the $x$ axis for
  $N=500$ (lower curves) and $N=1270$ (upper curves).
 The  solid lines are the results of our variational calculation.
The dotted  line are the  results obtained by using the Gaussian trial
wave function of Ref.\cite{baym}. 
\item[FIG. 6.]Aspect ratio in $^{7}$Li as a function of $N$. 

\end{itemize}
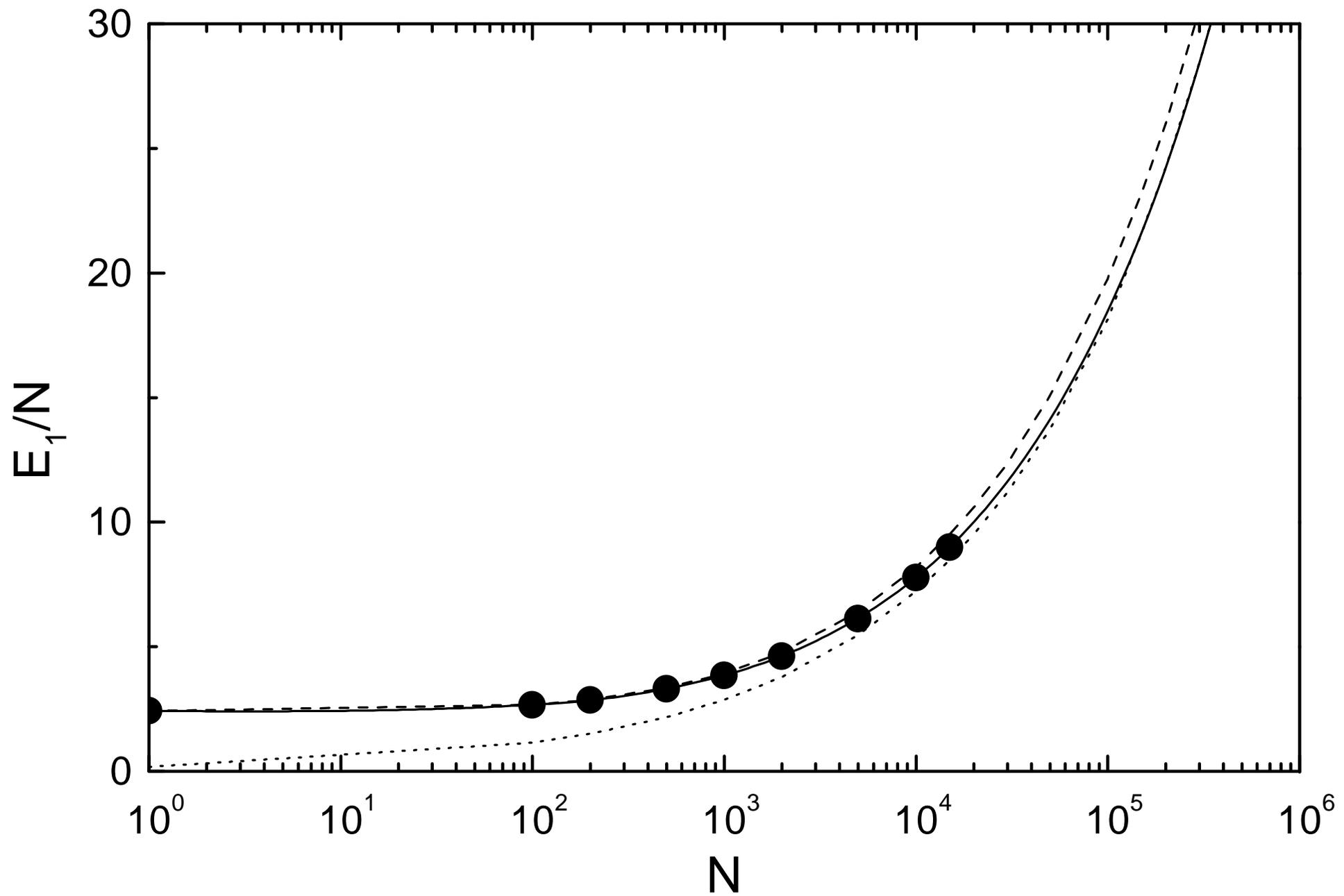
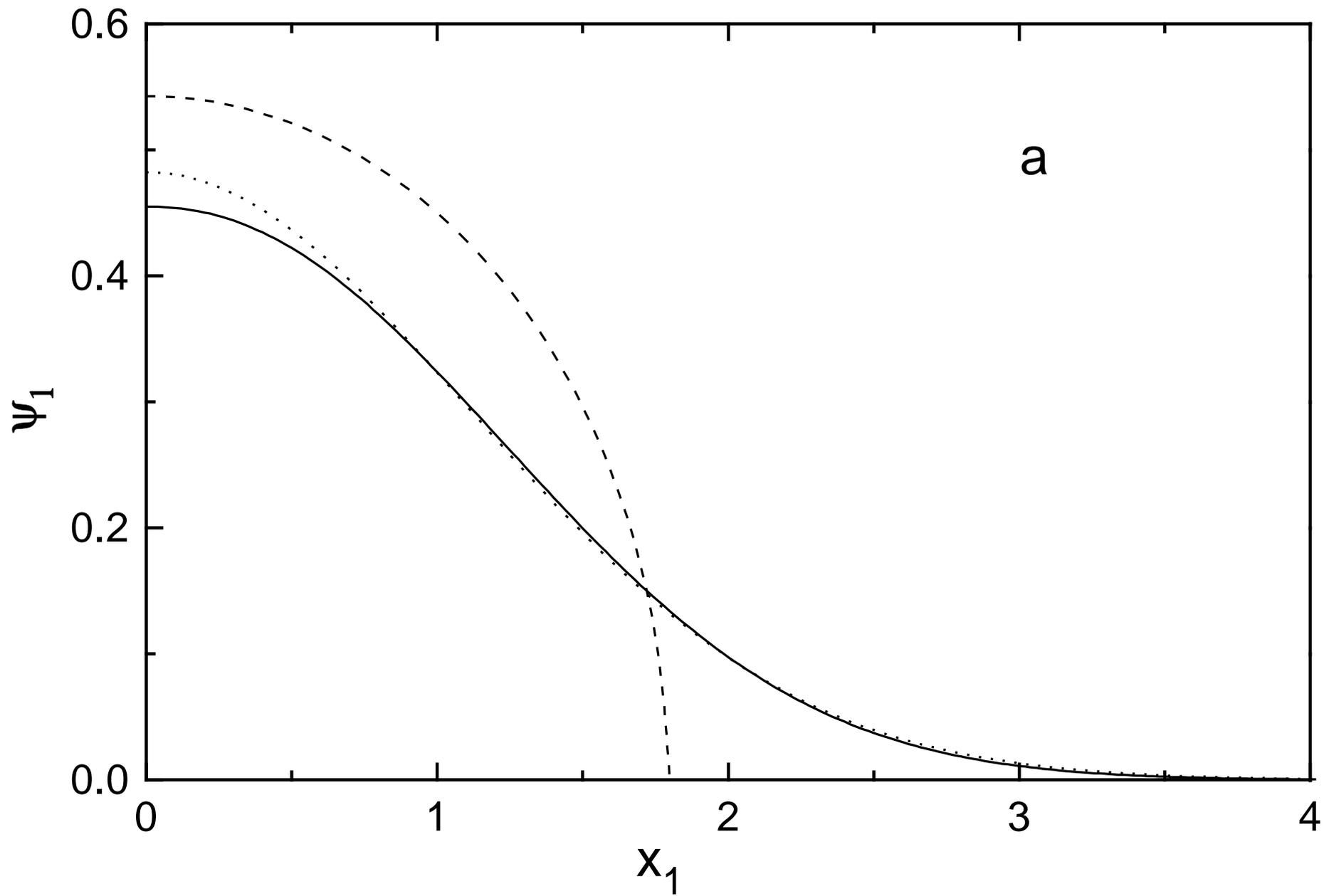
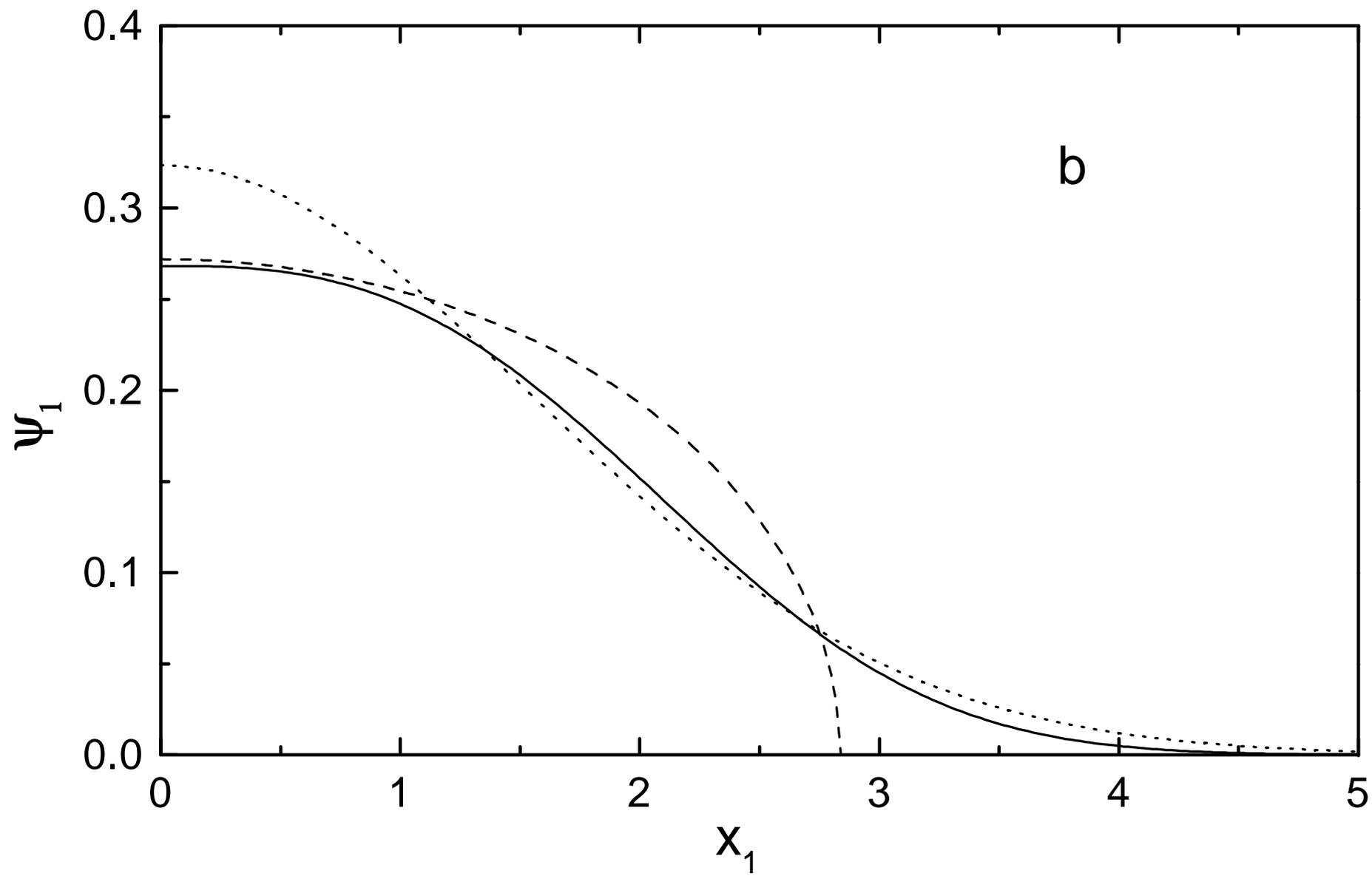
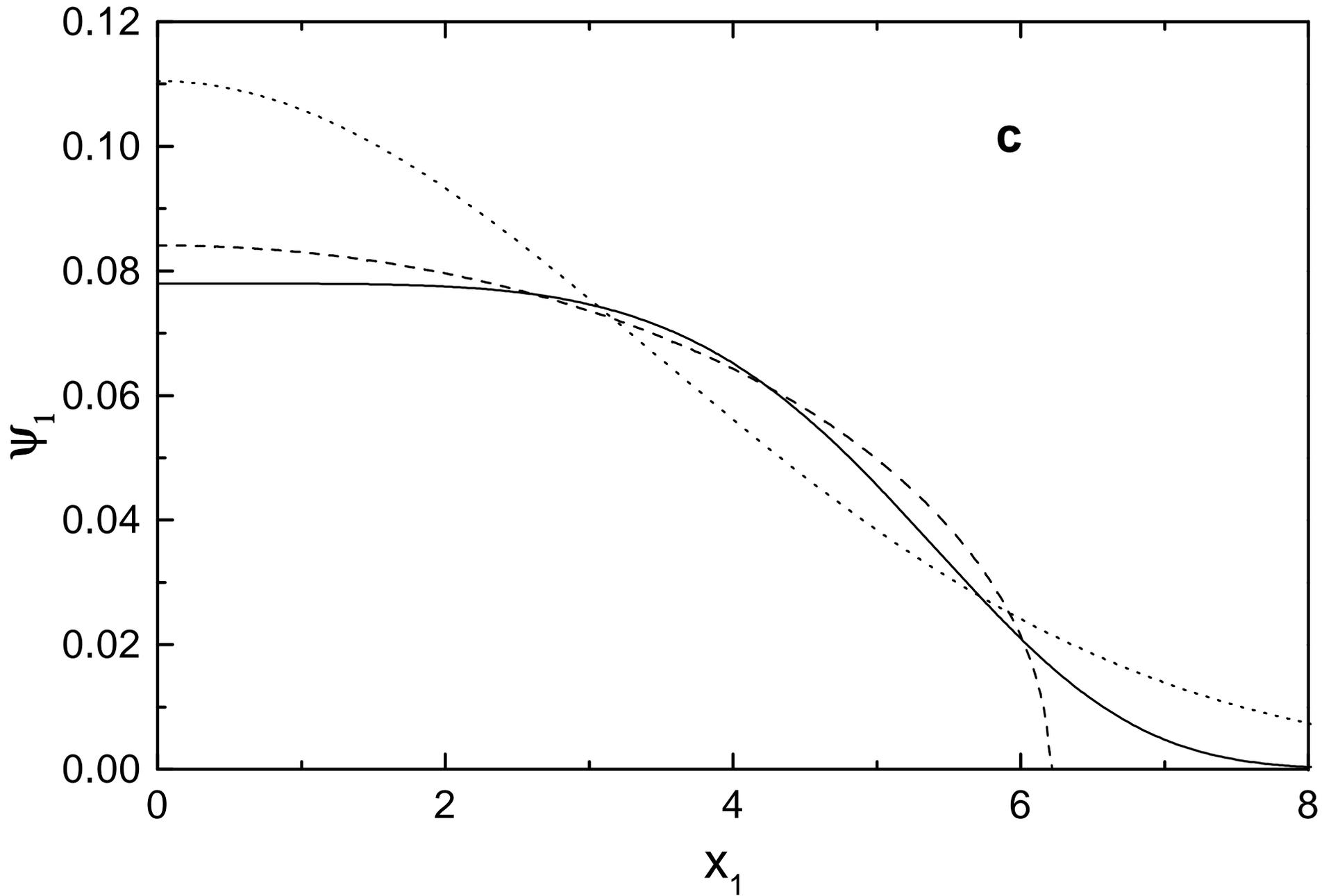
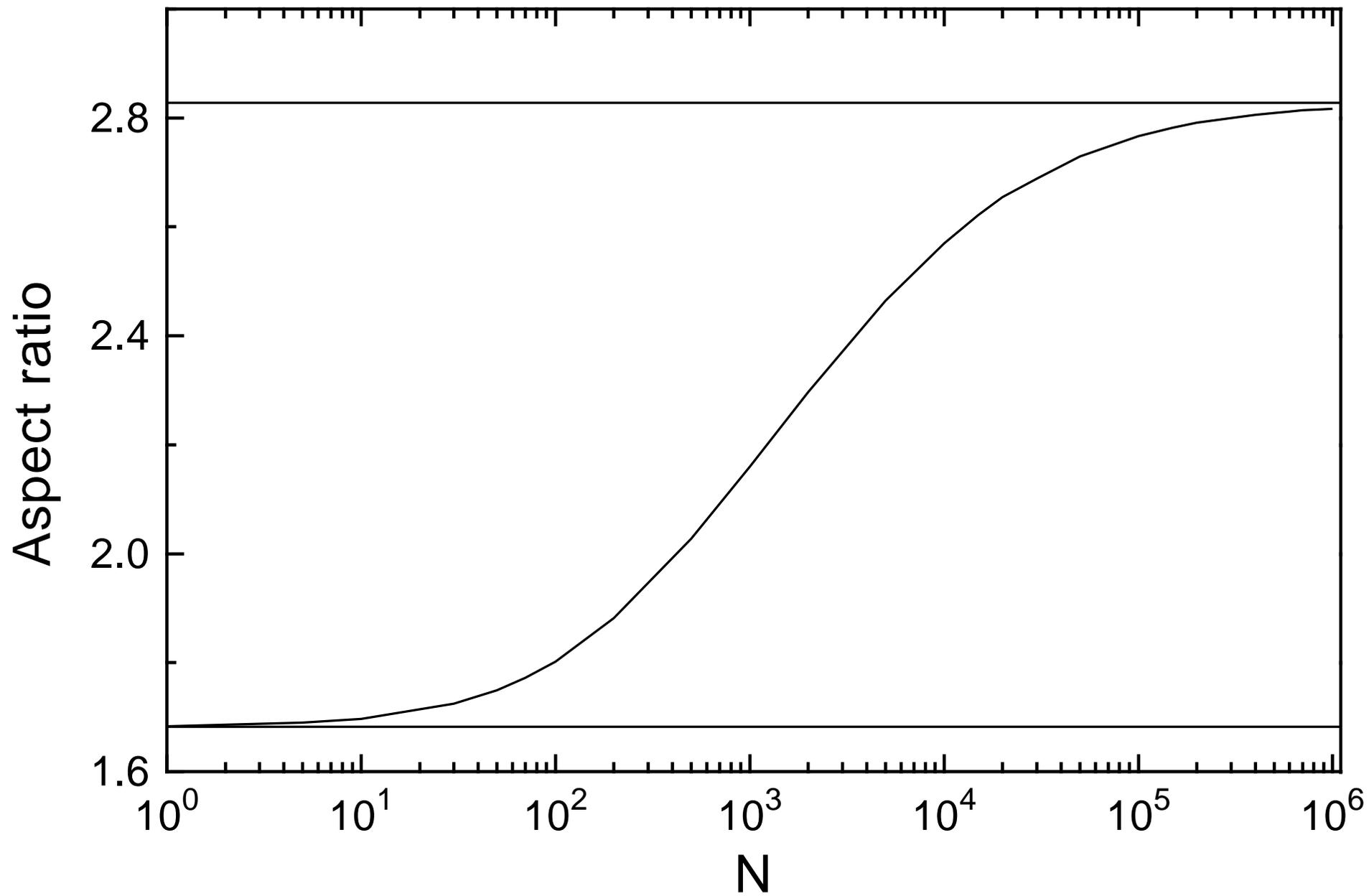
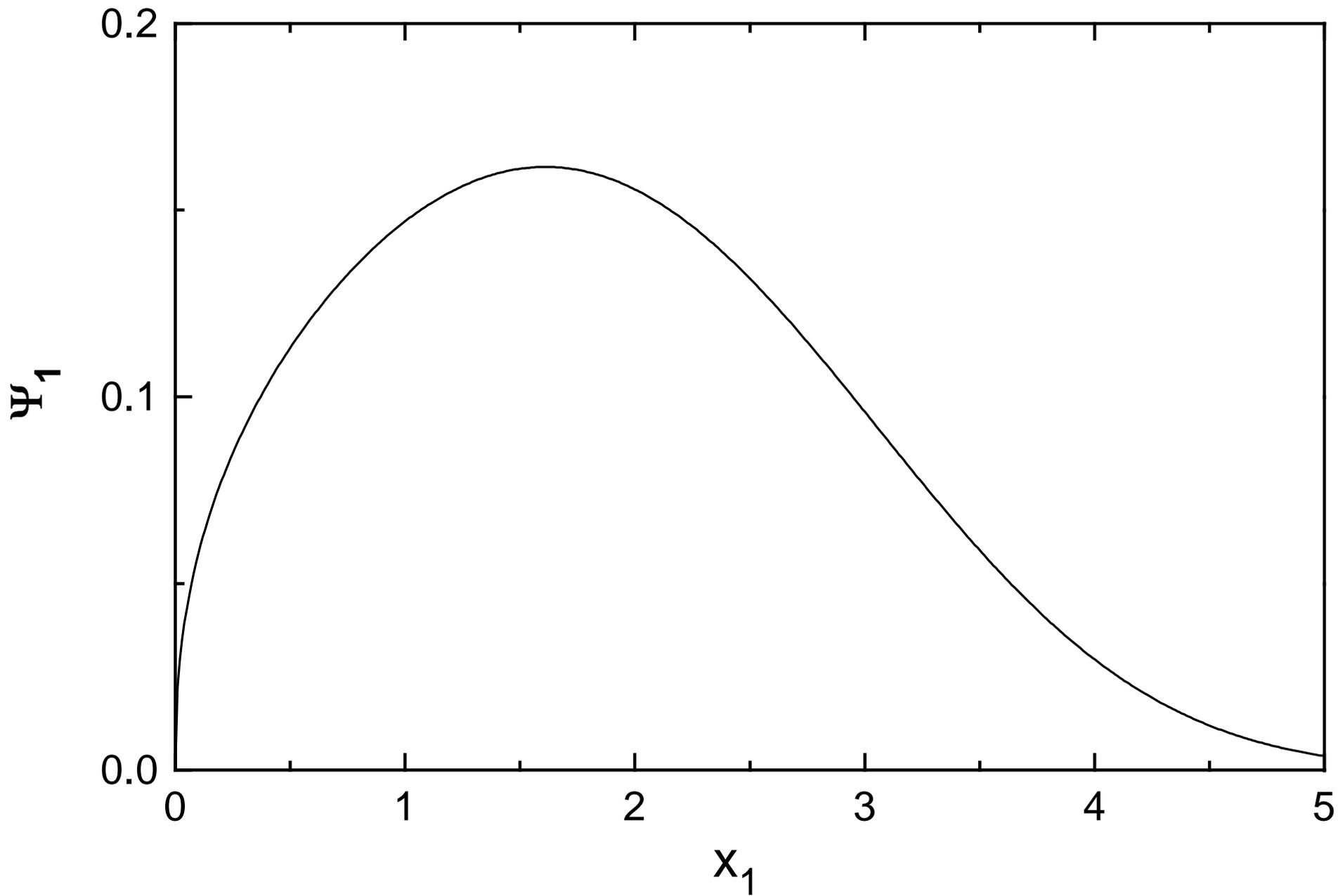
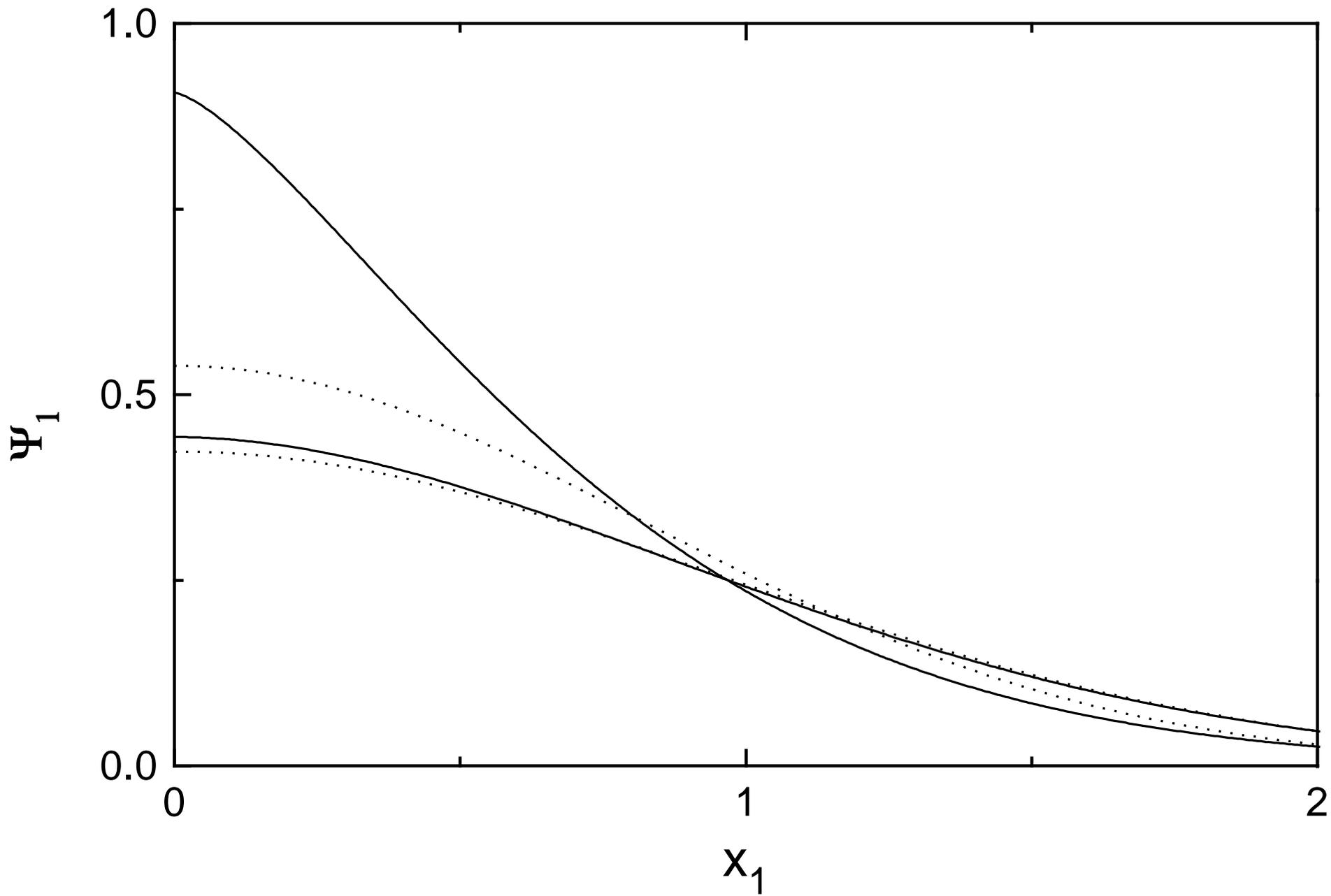
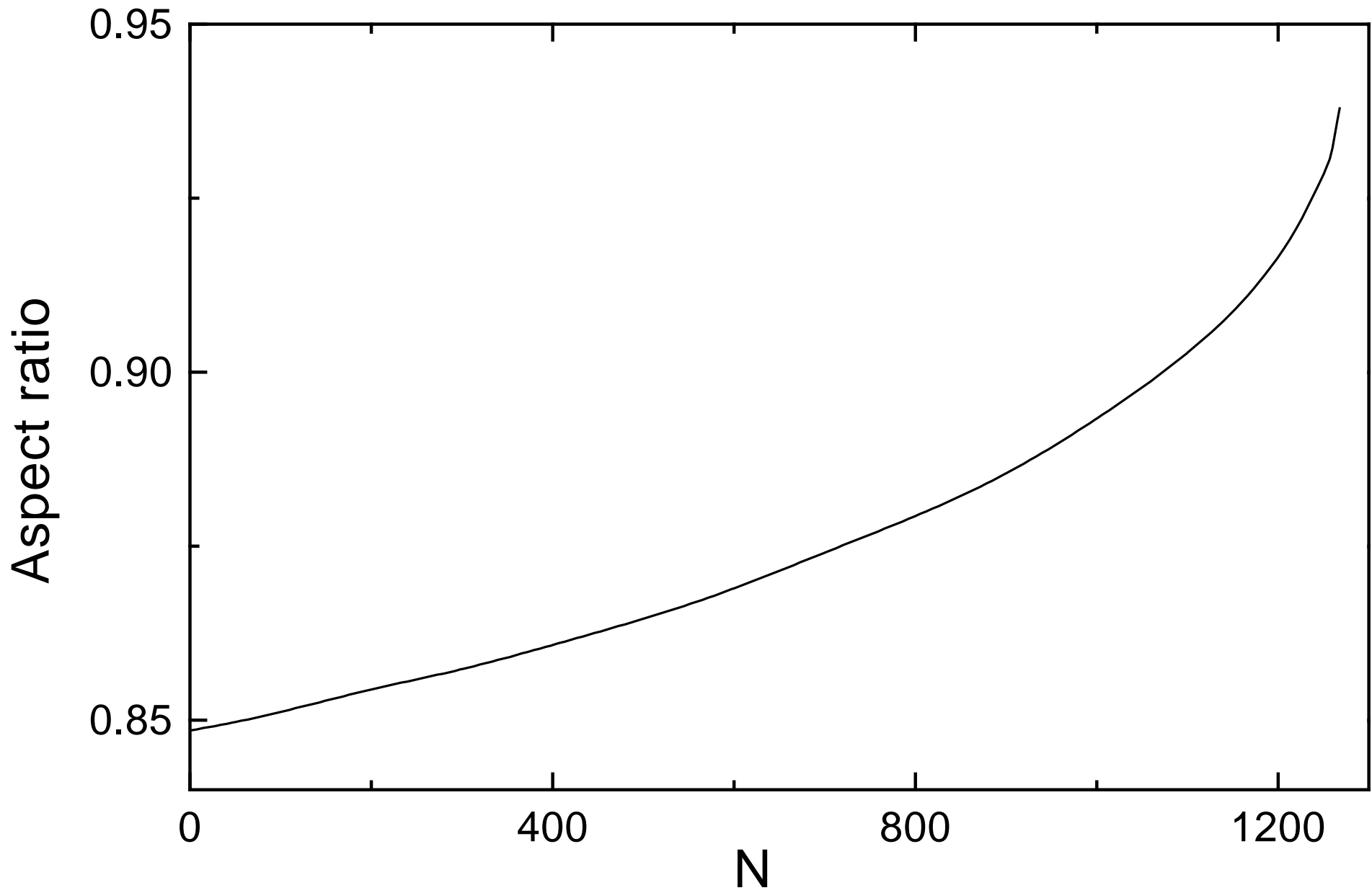
\end{document}